\documentstyle[12pt,epsf]{article}
\def\photonatomright{\begin{picture}(3,1.5)(0,0)
                                \put(0,-0.75){\tencircw \symbol{2}}
                                \put(1.5,-0.75){\tencircw \symbol{1}}
                                \put(1.5,0.75){\tencircw \symbol{3}}
                                \put(3,0.75){\tencircw \symbol{0}}
                      \end{picture}
                     }

\def\photonatomup{\begin{picture}(1.5,3)(0,0)
                             \put(-0.75,3){\tencircw \symbol{3}}
                             \put(-0.75,1.5){\tencircw \symbol{2}}
                             \put(0.75,1.5){\tencircw \symbol{0}}
                             \put(0.75,0){\tencircw \symbol{1}}
                   \end{picture}
                  }

\def\photonright{\begin{picture}(30,1.5)(0,0)
                     \multiput(0,0)(3,0){10}{\photonatomright}
                  \end{picture}
                 }

\def\photonrighthalf{\begin{picture}(30,1.5)(0,0)
                     \multiput(0,0)(3,0){5}{\photonatomright}
                  \end{picture}
                 }

\def\photonup{\begin{picture}(1.5,30)(0,0)
                  \multiput(0,0)(0,3){10}{\photonatomup}
               \end{picture}
              }
\def\photonuphalf{\begin{picture}(1.5,15)(0,0)
                      \multiput(0,0)(0,3){5}{\photonatomup}
                   \end{picture}
                  }

\def\fermionup{\begin{picture}(1,30)(0,0)
                     \put(0,0){\vector(0,1){15}}
                     \put(0,15){\line(0,1){15}}
               \end{picture}
              }
\def\fermionuphalf{\begin{picture}(1,15)(0,0)
                         \put(0,0){\vector(0,1){7.5}}
                         \put(0,7.5){\line(0,1){7.5}}
                   \end{picture}
                  }

\def\fermionull{\begin{picture}(30,15)(0,0)
                        \put(0,0){\vector(-2,1){15}}
                        \put(-15,7.5){\line(-2,1){15}}
                  \end{picture}
                 }
\def\fermionullhalf{\begin{picture}(15,7.5)(0,0)
                        \put(0,0){\vector(-2,1){7.5}}
                        \put(-7.5,3.75){\line(-2,1){7.5}}
                  \end{picture}
                 }
\def\fermionurr{\begin{picture}(30,15)(0,0)
                        \put(-30,-15){\vector(2,1){15}}
                        \put(-15,-7.5){\line(2,1){15}}
                  \end{picture}
                 }
\def\fermionurrhalf{\begin{picture}(15,7.5)(0,0)
                        \put(-15,-7.5){\vector(2,1){7.5}}
                        \put(-7.5,-3.75){\line(2,1){7.5}}
                  \end{picture}
                 }

\newenvironment{Feynman}[3]{\begin{center}
                            \setlength{\unitlength}{#3 mm}
                            \begin{picture}(#1)(#2)
                            \thicklines
                           }{\end{picture} \end{center}}
\newcommand {\oalf} {\mbox{${\cal O}(\alpha)$}}

\newcommand{\nn}{\noindent}

\newcommand{\bq}{\begin{equation}}
\newcommand{\eq}{\end{equation}}
\newcommand{\ba}{\begin{eqnarray}}
\newcommand{\ea}{\end{eqnarray}}

\newcommand{\mathrm}{\rm}
\newcommand{\mr}{\mathrm}
\newcommand{\ww}{\mbox{$\Gamma_W$}}
\newcommand{\wm}{\mbox{$M_W$}}
\newcommand{\nl}{ \nonumber \\}
\newcommand{\ltwo}{LEP~200}
\begin{document}
\thispagestyle{empty}
\onecolumn
\begin{flushleft}
DESY 93--035
\\
BI--TP 93/09
\\
March 1993
\\
Phys. Letters B308 (1993) 403
\\
revised July 1995
\end{flushleft}
\vspace*{2.00cm}
\nn
\hspace*{-5mm}
\begin{center}
{\Large  \bf
Off-shell $W$-pair production in $e^+ e^-$-annihilation
\vspace*{0.6cm}}
\\ {\Large \it -- Initial state radiation --}
\vspace*{2.0cm}
\end{center}
\nn
{\large
D. Bardin$\;^{1,2}$,
$\;$ M. Bilenky$\;^{1,2}$, $\;$
A.~Olchevski$\;^{2 \mr a} \; $
and
$\;$ T.~Riemann$\,^1$
}
\\
\vspace*{0.5cm}

\begin{normalsize}
\begin{tabbing}
$^1$  \=
DESY -- Institut f\"ur Hochenergiephysik
\\ \>
Platanenallee 6, D--15738 Zeuthen, Germany
\end{tabbing}
\begin{tabbing}
$^2$ \=
Bogoliubov Laboratory for Theoretical Physics, JINR
\\ \>
ul. Joliot-Curie 6, RU--141980 Dubna, Moscow Region, Russia
\end{tabbing}
\end{normalsize}

\vspace*{1.5cm}
\vfill
\thispagestyle{empty}
{\large
\centerline{
ABSTRACT}
\vspace*{.3cm}
\nn
\normalsize
With a current-splitting technique, we calculate the gauge-invariant
initial-state radiation to order \oalf\ with soft-photon exponentiation
for on- and off-shell $W$-pair production.
This result generalizes the convolution formula, which is known from the
description of the $Z$ resonance, to the case of the production of two
$W$-bosons.
After up to eightfold analytical integrations, a sufficiently smooth
integral over three invariant masses remains to be treated numerically.
Including the Coulomb
singularity, the largest corrections are covered.
We discuss the corrections in a large energy range up to
$\sqrt{s}$=1~TeV and draw numerical conclusions on their influence on the
$W$-mass determination at LEP~200.
\\ Table~1 and figure~5 are revised after correcting the analytical
formulae for the nonuniversal corrections.
\vspace*{.3cm}
\nn
\normalsize
} 
\vspace*{.5cm}

\bigskip
\vfill
\footnoterule
\nn
{\small
$^a$ Present address: CERN, Geneva, Switzerland
\begin{tabbing}
email: \= BARDINDY@CERNVM.CERN.CH, BILENKYM@VXCERN.CERN.CH,
\\      \> OLSHEVSK@VXCERN.CERN.CH, riemann@ifh.de
\end{tabbing}
}
\newpage

\section{Introduction}

The Standard Model of electroweak interactions allows the
calculation of the
width $\Gamma_W$~\cite{wdecay} and mass \wm\ of the $W$~boson;
the latter may be iterated from $\Delta r$~\cite{deltar}, the
electroweak correction to the muon decay constant.
For fixed \mbox{t-quark} and Higgs masses,
the predictions of the Standard Model are much more precise than the
available experimental data.

A direct measurement of $\Gamma_W$ and \wm\ will be one of the
main tasks of \ltwo~{\mbox{\cite{aachen,camilleri}}}.
This will be possible from a study of the reaction
\ba
e^+ e^- \rightarrow (W^+ W^-) \rightarrow 4 f,
\label{eqborn}
\ea
which is accompanied by the emission of photons~\cite{aeppli} and gluons,
\ba
e^+ e^-     
\rightarrow 4 f + n_1 \gamma + n_2 g.
\label{eqrc}
\ea
The photons may be emitted by the initial and intermediate states;
and both photons and gluons by final state particles.
With a centre-of-mass energy at \ltwo\ of
$\sqrt{s} \approx 190$ GeV, the $W$-pair production proceeds near
the threshold.
Thus, the finite width effects in the off-shell production must be taken
into account.
And last but not least, virtual electroweak corrections have to be
inserted
properly.

Here, we will
concentrate on the treatment of {\em Initial State
Radiation} (ISR)~\footnote
{
In another article~\cite{isr2}, we classify and study several background
reactions with the same signature as~(\ref{eqborn}), but different intermediate
states. They have to be added in order to ensure gauge invariance for the
process.
}.
The aim is to obtain (semi-)analytical expressions for the total cross section
at \ltwo, but also at higher energies~\cite{ws500a,ws500b}.
In section~2, we introduce the Born cross section and define the notations.
In section~3, photonic bremsstrahlung will be faced.
Since the $W$-boson is electrically charged, we have to derive properly a
gauge-invariant definition of ISR.
Section~4 contains the cross section formulae with contributions from ISR.
For the applications at LEP~200, we discuss in section~5 a possible
estimate of the Coulomb singularity.
Numerical results and some conclusions
both for LEP~200 and for a wide energy range
are presented in section~6.
%
\section{The Born cross section}
%
The main contributions to reaction~(\ref{eqborn}) are
shown in figure~1: two {\tt crayfish} diagrams and one {\tt crab} diagram.
We use the unitary gauge; in a non-unitary gauge,
one should have to take into account additional diagrams.
The cross section is well described by a
twofold convolution of a hard-scattering off-shell cross section~\cite{muta}:
\bq
\sigma_{\mr B}^{\mr{off}}(s)
=
\int_0^s ds_1 \, \rho(s_1)
\int_0^{(\sqrt{s} - \sqrt{s_1})^2} ds_2 \, \rho(s_2) \,
\sigma_0(s;s_1,s_2),
\label{sigmab}
\eq
\ba
\rho(s_i)
=
\frac{1}{\pi}
\frac {\sqrt{s_i} \, \ww (s_i) }
      {|s_i - M_W^2 + i \sqrt{s_i} \, \ww (s_i) |^2} \times {\mr{BR}}(i),
\label{rho}
\ea
\bq
\ww (s_i)
=
 \sum_f \frac{G_{\mu}\, \wm ^2} {6 \pi \sqrt{2}}
 \sqrt{s_i}.
\label{gwoff}
\eq

\begin{minipage}[tbh]{7.8cm}{
\begin{center}
\begin{Feynman}{75,60}{-4,0}{0.8}
%
\put(30,30){\fermionurr}
\put(30,30){\fermionull}
\put(30,30){\photonright}
\put(30,30){\circle*{1.5}}
\put(60,30){\circle*{1.5}}
\put(60,15){\circle*{1.5}}
\put(60,45){\circle*{1.5}}
\put(75,7.5){\fermionullhalf}
\put(75,22.5){\fermionurrhalf}
\put(60,15){\photonuphalf}
\put(60,30){\photonuphalf}
\put(75,37.5){\fermionullhalf}
\put(75,52.5){\fermionurrhalf}
\small
\put(-07,48){$e^+(k_2)$}  
\put(-07,09){$e^-(k_1)$}
\put(36,24){$\gamma, Z$}
\put(50,36.5){$W^+$}  
\put(50,18.0){$W^-$}
\put(77,06){${\bar f}_1^u(p_2)$}  
\put(77,22){$       f_1^d(p_1)$}
\put(77,36){${\bar f}_2^d(p_4)$}  
\put(77,52){$       f_2^u(p_3)$}
\normalsize
\end{Feynman}
\end{center}
}\end{minipage}
\begin{minipage}[tbh]{7.8cm} {
\begin{center}
\begin{Feynman}{75,60}{06,0}{0.8}
%
\put(30,15){\fermionurrhalf}
\put(30,45){\fermionullhalf}
\put(30,15){\fermionup}
\put(30,45){\photonright}
\put(30,15){\photonright}
\put(30,45){\circle*{1.5}}
\put(30,15){\circle*{1.5}}
\put(60,15){\circle*{1.5}}
\put(60,45){\circle*{1.5}}
\put(75,7.5){\fermionullhalf}
\put(75,22.5){\fermionurrhalf}
\put(75,37.5){\fermionullhalf}
\put(75,52.5){\fermionurrhalf}
\small
\put(32,28){$\nu(q)$}
\put(38,49){$W^+(w_2)$}  
\put(38,09){$W^-(w_1)$}
\normalsize
\end{Feynman}
\end{center}
}\end{minipage}
\vspace*{.5cm}

\begin{center}
\noindent
{Figure 1:} {\it
The Feynman diagrams for off-shell $W$-pair production:}
{\tt crayfish} {\it and} {\tt crab}.
\vspace*{.5cm}
\end{center}

The off-shell width $\ww (s_i)$ contains a sum over all open
fermion-decay channels $f$ at energy $\sqrt{s_i}$, and BR($i$) is the
corresponding branching fraction.
We have rederived the results of~\cite{muta} and refer for details to that
article.
For later use, we split the cross section into three pieces:
\ba
\sigma_0 =
\sigma_0^{\mr s}    +
\sigma_0^{\mr{st}}   +
\sigma_0^{\mr t}.
\label{sigsstt}
\ea


\section{Photonic corrections to $W$-pair production}
%
In the description of the complicated scattering process~(\ref{eqrc}),
it
seems to be desirable to find gauge-invariant subsets of Feynman diagrams.
Further, one would like to separate in a reasonable way the terms related to
bremsstrahlung from the genuine electroweak virtual corrections.

A procedure to disentangle final-state real photon emission from the rest of
the process~(\ref{eqrc}) has been proposed in~\cite{kleiss1}.
A corresponding separation of virtual photonic corrections should be performed
in parallel.
In the presence of $W$ exchange, these do not form a gauge-invariant subset of
diagrams.
Being mainly
interested in real bremsstrahlung, one could decide to combine it only with
the singular parts of the virtual diagrams~\cite{cchera}.
The singular parts are then uniquely defined up to an additive constant and
will compensate the singularities of the real bremsstrahlung.

These few remarks should indicate that there is much work to be done if one
intents to perform a systematic, semi-analytical study of
reaction~(\ref{eqrc}).
For a Monte-Carlo approach, we refer to~\cite{aeppli}.

In the following, we will concentrate on ISR. Thus, we avoid to be faced with
the full complexity of the problems, and at the same time we will cover the
bulk of the numerically important corrections.
The initial-state radiation offers no problems in the case of radiation
from the {\tt crayfish} diagrams. Here, the initial charges are separated by
the
exchange of neutral gauge bosons from the rest of the diagrams, and ISR
is well-defined.

%
\subsection{The $t$-channel ISR problem -- splitting the neutrino current}
%

The {\tt crab} diagram contains a neutrino exchange in the $t$ channel.
The photonic corrections to the initial state are shown in figure~2.

\begin{minipage}[tbhp]{15.cm} {
\begin{center}
\begin{Feynman}{150,60}{-22,0}{0.933}
\small
\put(-09,52.5){$e^+(k_2)$}  
\put(-09,06.5){$e^-(k_1)$}
\put(30,49){$\gamma(p)$}
\normalsize
\put(20,45){\line(-2,1){15.00}}
\put(08.75,50.625){\vector(-2,1){1}}  
\put(16.25,46.875){\vector(-2,1){1}}
\put(20,15){\fermionurrhalf}
\put(20,15){\fermionup}
\put(12.5,48.75){\photonrighthalf}
\put(20,45){\circle*{1.5}}
\put(20,15){\circle*{1.5}}
\put(12.5,48.75){\circle*{1.5}}
%
\put(60,45){\fermionullhalf}
\put(60,15){\fermionup}
\put(52.5,11.25){\photonrighthalf}
\put(60,45){\circle*{1.5}}
\put(60,15){\circle*{1.5}}
\put(52.5,11.25){\circle*{1.5}}
\put(45,7.5){\line(2,1){15.00}}
\put(48.75,9.375){\vector(2,1){1}}  
\put(57.25,13.625){\vector(2,1){1}}
%
\put(100,15){\fermionurrhalf}
\put(115,22.5){\fermionurrhalf}
\put(100,45){\fermionullhalf}
\put(115,37.5){\fermionullhalf}
\put(115,22.5){\fermionuphalf}
\put(100,15){\photonup}
\put(100,45){\circle*{1.5}}
\put(100,15){\circle*{1.5}}
\put(115,22.5){\circle*{1.5}}
\put(115,37.5){\circle*{1.5}}
%
\end{Feynman}
\end{center}
}\end{minipage}
\vspace*{.3cm}

\begin{center}
\noindent
{Figure 2:} {\it
The usual initial-state photonic corrections to the} {\tt crab} {\it
diagram.
}
\vspace*{.5cm}
\end{center}

The bremsstrahlung
   may be described by the following electromagnetic current:
\ba
{\cal J}_{\mu, \alpha  \beta}^{\mr{br}}
&=&
Q_e {\bar u}(k_2)
\left[
\gamma_{\alpha}
\frac{{\hat w}_2-{\hat k}_2}{(w_2-k_2)^2}
\gamma_{\beta}
\frac{2 k_1^{\mu}-{\hat p} \gamma_{\mu}}{-2k_1p}
-
\frac{2 k_2^{\mu}- \gamma_{\mu}{\hat p}}{-2k_2p}
\gamma_{\alpha}
\frac{-{\hat w}_1+{\hat k}_1}{(w_1-k_1)^2}
\gamma_{\beta}
\right]
\left(1+\gamma_5\right)u(k_1),
\nl
\label{crab1}
\ea
where the $\{ k_1,k_2,p,w_1,w_2\}$ are the four-momenta of
$\{e^-,e^+,\gamma,W^-,W^+\}$, respectively.
This current is not conserved,
$p^{\mu}{\cal J}_{\mu, \alpha \beta}^{\mr{ br}}
\neq 0$,
and gauge-invariance is violated.
Nevertheless,
one may construct a conserved current also for the {\tt crab}
diagram.
Intuitively, it is evident that one has to allow the
flow of charge {\em within} the initial state;
a charge $Q_e$  has to be associated to the neutrino propagator.
This, in turn, enables the neutrino to radiate, see figure~3.
The resulting auxiliary current,
\ba
{\cal J}_{\mu, \alpha \beta}^{\mr{aux}}
&=&
Q_e {\bar u}(k_2)
\left[
\gamma_{\alpha}
\frac{{\hat w}_2-{\hat k}_2}{(w_2-k_2)^2}
\gamma_{\mu}
\frac{-{\hat w}_1+{\hat k}_1}{(w_1-k_1)^2}
\gamma_{\beta}
\right]
\left(1+\gamma_5\right)u(k_1),
\label{cursplit}
\ea
ensures current conservation in the initial state and thus makes the      d
combined diagrams gauge-invariant.
In fact, the net initial-state current,
\ba
{\cal J}_{\mu, \alpha \beta}^{\mr {ini}}
&\equiv&
{\cal J}_{\mu, \alpha \beta}^{\mr{br}}
+
{\cal J}_{\mu, \alpha \beta}^{\mr{aux}},
\label{crab2}
\ea
fulfills current conservation,
\ba
p^{\mu}{\cal J}_{\mu, \alpha \beta}^{\mr{ ini}}
= 0.
\label{null}
\ea
%
In order to compensate for the extra piece in the matrix element, one has to
add also a diagram with flow of charge $-Q_e$ through the neutrino propagator,
which in its turn
has to be combined with the charge flow through the $W$ bosons
to build also a continuous electric charge flow
-- but now as part of the {\em intermediate state} of the process
(or, in case of on-shell $W$-pair production, of the final state).
In effect, the electrically neutral neutrino has been
split into two oppositely flowing, equal charges~\footnote
{
          For quarks, the auxiliary diagrams from the
          `charged' neutrino are naturally present, and by no
          means auxiliary. For an application of the CST in $ep$ scattering
          at HERA, see~\cite{cchera}.
}.
%

\begin{minipage}[tbhp]{15.cm} {
\begin{center}
\begin{Feynman}{150,60}{0,0}{0.933}

\put(20,15){\fermionurrhalf}
\put(20,45){\fermionullhalf}
\put(20,30){\fermionuphalf}
\put(20,15){\fermionuphalf}
\put(20,30){\photonrighthalf}
\put(20,45){\circle*{1.5}}
\put(20,15){\circle*{1.5}}
\put(20,30){\circle*{1.5}}
\put(60,15){\fermionurrhalf}
\put(60,45){\fermionullhalf}
\put(60,15){\fermionuphalf}
\put(60,30){\photonuphalf}
\put(60,37.5){\oval(15,15)[r]}
\put(65.4,42.8){\vector(-1,1){1}}  
\put(65.45,32.2){\vector(1,1){1}}  
\put(60,45){\circle*{1.5}}
\put(60,15){\circle*{1.5}}
\put(60,30){\circle*{1.5}}
\put(67.50,37.5){\circle*{1.5}}
\put(100,15){\fermionurrhalf}
\put(100,45){\fermionullhalf}
\put(100,30){\fermionuphalf}
\put(100,15){\photonuphalf}
\put(100,22.5){\oval(15,15)[r]}
\put(105.35,27.8){\vector(-1,1){1}}  
\put(105.35,17.2){\vector(1,1){1}}  
\put(100,45){\circle*{1.5}}
\put(100,15){\circle*{1.5}}
\put(100,30){\circle*{1.5}}
\put(107.50,22.5){\circle*{1.5}}
\put(140,15){\fermionurrhalf}
\put(140,45){\fermionullhalf}
\put(140,15){\line(0,1){7.5}}
\put(140,18.75){\vector(0,1){1}}  
\put(140,41.25){\vector(0,1){1}}  
\put(140,37.5){\line(0,1){7.5}}
\put(140,22.5){\photonuphalf}
\put(140,30){\oval(15,15)[r]}
\put(147.48,30){\vector(0,1){1}}  
\put(140,45){\circle*{1.5}}
\put(140,15){\circle*{1.5}}
\put(140,37.5){\circle*{1.5}}
\put(140,22.5){\circle*{1.5}}
\end{Feynman}
\end{center}
}\end{minipage}
\vspace*{.3cm}

\begin{center}
\noindent
{Figure 3:} {\it
The auxiliary initial-state photonic corrections to the} {\tt crab} {\it
diagram.
}
\vspace*{0.3cm}
\end{center}

For this reason, we call this method of restoring gauge invariance for
ISR the {\em Current Splitting Technique} (CST).

The introduction of an additional bremsstrahlung diagram to ISR has
a further consequence: one must take into account also additional
virtual corrections.
These are also shown in figure~3.
They, together with the corresponding counter terms, will
ensure the compensation of the infrared divergencies of the photon radiation
from the beam particles.
The counter terms also cancel the UV-divergencies of the newly
introduced virtual corrections.
The additional radiation from the neutrino line is free of any infrared
divergency, since the radiating particle is off-shell.
%
%
%
\section{Initial-state radiation}
%
For the $s$-channel {\tt crayfish} diagram we rederived the ISR correction
factor, which is well-known from $Z$-resonance calculations;
see e.g.~\cite{berends} and references cited therein.
The corresponding cross section is described by a threefold convolution:
\bq
\frac  {d^2 \sigma_{\mr s}}
       {ds_1 ds_2}
=
\int_{
(\sqrt{s_1}+\sqrt{s_2})^2
} ^s \frac{ds'}{s}
\left[ \beta_e v^{\beta_e - 1} (1+ \bar S) + {\bar H} \right]
\, \rho(s_1) \, \rho(s_2) \,
\sigma_0^{\mr s}(s';s_1,s_2).
\label{qed-s}
\eq
Here, it is $v=1-s'/s$ and
the soft plus virtual photon part ${\bar S}(s)$ and hard part
${\bar H} (s,s')$ are:
\ba
{\bar S}(s) = \frac{\alpha}{\pi} \left[ \frac{\pi^2}{3} - \frac{1}{2}
\right]+ \frac{3}{4} \beta_e + {\cal O}(\alpha^2) ,
\hspace{.7cm}
{\bar H}(s,s') = - \frac{1}{2} \left(1+\frac{s'}{s}\right)
\beta_e
+ {\cal O}(\alpha^2) ,
\label{betae}
\ea
and $\beta_e=2 \alpha/\pi [ \ln (s/m_e^2) - 1 ]$.

%
\subsection{The ISR connected with the {\tt crab} diagram}
%
For a treatment of single-photon bremsstrahlung, one has to integrate a
tenfold distribution.
The degrees of freedom of the four-fermion final state have been chosen to be
two pairs of angles in the rest systems of the two fermion pairs.
These integrations are trivially performed with the
method of tensor integration (see e.g.~\cite{landau}).
As further integration variables, we took the cosine of the photon angle
$\theta_{\gamma}$ in the centre-of-mass system and the two angles
$\phi_{W},\theta_{W}$ of the virtually produced
$W^-$-boson in the recoil system, i.e. with respect to the photon axis.
It is the integration over these three angles which demands some effort,
and which leads to the tremendous complications in the case of the
{\tt crab} diagram with a neutrino exchange in the $t$ channel.

Details of the calculation will be published elsewhere.
The algebraic manipulations have been performed with the aid of the
programs for algebraic manipulations {\tt SCHOONSCHIP} and
{\tt FORM}~\cite{schoonschip}.
We finally arrived at the following cross section:
\bq
\frac {d^2 \sigma_{\mr{a}}} {ds_1 ds_2}
=
\int_{(\sqrt{s_1}+\sqrt{s_2})^2}^s \frac{ds'}{s}
\rho(s_1) \, \rho(s_2)
\left[ \beta_e v^{\beta_e - 1} {\cal S}_{\mr a}
+{\cal H}_{\mr a}
\right],
\label{qed-t}
\eq
\ba
{\cal S}_{\mr a}(s,s';s_1,s_2)
&=&
\left[ 1 + {\bar S}(s) \right] \sigma_0^{\mr{a}}(s';s_1,s_2)
+ \sigma^{\mr a}_{\hat S}(s';s_1,s_2),
\label{hats}
\\
{\cal H}_{\mr a}(s,s';s_1,s_2)
&=&
{\bar H}(s,s') \sigma_0^{\mr a}(s';s_1,s_2)
+ \sigma^{\mr a}_{\hat H}(s,s';s_1,s_2).
\label{hath}
\ea
Here, a=${\mr{st,t}}$ denote the $st$-interference and $t$-channel
contributions, respectively.
The functions
${\cal S}_{\mr a}(s,s';s_1,s_2)$ and
${\cal H}_{\mr a}(s,s';s_1,s_2)$
contain extra cross-section pieces with deviations
from the structure of the $s$-channel case. As already mentioned, they are
{\em nonuniversal} in the sense that they differ for a=st,t.
Further, they are {\em not factorizing}, i.e. they do not contain the off-shell
Born cross section as an explicit factor.
A third feature of them is a {\em screening} property: compared to the Born
cross section, they have a damping overall factor,
\ba
\sigma^{\mr {st,t}}_{{\hat S},{\hat H}}(s';s_1,s_2)
\sim
\frac{s_1 s_2}{s^2},
\label{scree}
\ea
which suppresses potential mass singularities, which otherwise would be
generated by new kinematic logarithms compared to the Born cross section.
The non-universality of the additional corrections could, in principle,
spoil the {\em gauge cancellations}~\cite{buras} between the three
diagrams of figure~1.
{}From the analytical structure of the nonuniversal terms,
such a cancellation may not be seen;
the $\sigma^{\mr {a}}_{{\hat S},{\hat H}}(s';s_1,s_2)$
are complicated functions, as one may expect after seven to eight integrations.
The property~(\ref{scree}) is welcome in this respect, but it could be
compensated at least partly by the integrations over invariant masses;
numerically, we have observed that the net cross section behaves
properly
at extremely high energies.
A consequence of the auxiliary current is that our formulae for ISR
contain some contributions  which in other calculations  are considered
a part of the {\it intermediate or final-state radiation}.

Besides rational functions and logarithms of $s,s',s_1,s_2$, they contain
dilogarithms and, in case of the virtual corrections, trilogarithms.
We will publish the explicit expressions for the cross section elsewhere.
Despite the fact that they are relatively lengthy, we would like to stress
that they are much more compact than they would be without the
introduction of the auxiliary current~(\ref{cursplit}).
It is of importance for a numerical handling
that the $\sigma_{\hat S}^{\mr a}$ and $\sigma_{\hat H}^{\mr a}$ are small
in the \ltwo\ energy region.
This allows us to include soft-photon exponentiation and ${\cal O}(\alpha^2)$
leading-logarithmic corrections into the $st$- and $t$-channel universal terms,
and to treat the nonuniversal rest to order \oalf.

At the beginning of this section  we
reproduced, in the context of off-shell $W$-pair production,
a general result for $s$-channel ISR: the radiator functions
${\bar S}$ and ${\bar H}$,
which describe the creation of an intermediate vector boson $V^{*}$
together with a photon,
$e^+ e^- \rightarrow V^{*}(s') \gamma_{\mr{ ini}}$.
The corresponding $t$-channel contribution with two virtual vector bosons,
\ba
e^+ e^- \rightarrow V_1^*(s_1)  V_2^*(s_2) \gamma_{\mr{ ini}}
,
\label{vv}
\ea
and its interference with the $s$ channel,
is characterized mainly by the same, universal functions
${\bar S}$ and ${\bar H}$,
but also by some deviations ${\hat S}$ and ${\hat H}$,
due to the more involved kinematics.
It was explicitly assumed that $V_1^*$ and $V_2^*$ have no common
final-state interactions; such interactions would destroy the anticipated
kinematical situation.

%
\section{The Coulomb singularity}
%
In the threshold region, there is another type of photonic corrections
besides the ISR, which is potentially large: the
Coulomb singularity~\cite{coulomb}.
For off-shell $W$-pair production, it originates
in the $s$ channel from the insertion
of a virtual photon to the $\gamma W^+W^-$ and  $ZW^+W^-$ vertices;
in the $t$ channel, it is due to the $\nu W^+ W^- \gamma$ box diagram.
It is not completely clear to us how to take the Coulomb singularity into
account numerically, apart from the fact that it should appear as a universal,
positive factor in the net cross section, which takes into account the
off-shellness of the process and should
approach the known on-shell correction,
\bq
\sigma_C
= \sigma \left[1 +  C(s';s_1,s_2, M_W, \Gamma_W)\right],
\label{sc}
\eq
\bq
C(s';s_1,s_2, M_W, \Gamma_W \rightarrow 0)
\stackrel{s_i \rightarrow M_W}{\longrightarrow}
\frac{ \pi \alpha}{2\beta(s')},
\hspace{.7cm}
\beta(s') = \sqrt{1-4M_W^2/s'} .
\label{C}
\eq
With a Feynman-diagram calculation one may find the scalar one-loop
function, in which the Coulomb singularity is located.
Unfortunately, an evaluation of it without a knowledge of the kinematic
behaviour of the imaginary part of the $W$
propagator is impossible. A crude estimate, assuming for the
$W$-propagators~(\ref{rho}) under the loop integral that
$i \sqrt{s_i} \, \Gamma_W(s_i) \approx i s_i \Gamma_W / M_W$, yields for the
leading term in the small-width approximation:
\bq
C(s';s_1,s_2, M_W, \Gamma_W)
 =
\left(1 - 3 \frac{\Gamma_W^2}{M_W^2}\right)
\frac{\pi \alpha}{2{\bar \beta}(s';s_1,s_2)},
\hspace{0.7cm}
{\bar \beta} (s';s_1,s_2)
 =
\frac{1}{s'} \sqrt{\lambda(s';s_1,s_2)}.
\label{barbeta}
\eq
Here, $\lambda$ is the usual kinematic $\lambda$-function.

A            different approach  which is based on a non-relativistic
treatment of the threshold region  has been developed in~\cite{khoze1,khoze2}.
For details we refer to the original literature.
%
\section{Numerical results}
%
%
In the discussion of numerical results, we will concentrate
on two observables,
which may be used for a precise determination of the
$W$~mass at \ltwo~\cite{camilleri}. 
The calculations have been performed with the aid of the
FORTRAN program {\tt GENTLE}~\cite{gentle}.
The corrected numerical values have been produced with version {\tt
CC11} of July 1995.
We have used the following input parameter definitions, which slightly
deviate from what was used in 1993:
\ba
G_{\mu} &=& 1.16639 \times 10^{-5} \, GeV^{-2},
\nl
\alpha \equiv \alpha(2M_W) &=& 1/128.07,
\nl
s_W^2 \equiv s_W^{2,eff} &=& \frac{\pi \alpha}{\sqrt{2} G_{\mu} M_W^2},
\nl
M_Z &=& 91.1888 \, GeV,
\nl
\Gamma_Z &=& 2.49747,
\nl
M_W &=& 80.230 \, GeV,
\nl
\Gamma_W &=& \frac{9 G_{\mu} M_W^3}{6\pi\sqrt{2}}.
\ea

\medskip

We will discuss the radiative corrections both for the cases of on-shell and
off-shell $W$-pair production.
Our `minimal' radiative corrections are the universal corrections introduced
in~(\ref{betae}), but neglecting the indicated terms of the order of ${\cal
O}(\alpha^2)$~\cite{berends}.
Additionally, we may take into account:
\begin{itemize}
\item[(i)] the universal ${\cal O}(\alpha^2)$ terms in~(\ref{betae})
           $[U_2$=0,1$]$;
\item[(ii)] the nonuniversal cross sections $\sigma_{\hat S}$ and
            $\sigma_{\hat H}$ in~(\ref{hats}) and~(\ref{hath}),
            which are the main
            result of this study $[{\hat S}$=0,1; ${\hat H}$=0,1$]$;
\item[(iii)] the estimate of the Coulomb singularity~(\ref{barbeta})
            $[C$=0,1$]$.
\end{itemize}
%
\subsection{The $W$-pair excitation curve}
%
We should like to start the discussion with a few comments on
figure~\ref{fig4}, where
we show the cross sections
$\sigma_{\mr B}^{\mr{on}}$ and $\sigma_{\mr B}^{\mr{ off}}$
 for Born on- and off-shell $W$-pair production, and
also $\sigma_{\mr T}^{\mr{ off}}$
for the off-shell production with universal ISR ($U_2$=1,
${\hat S}$=${\hat H}$=$C$=0).
Compared to $\sigma_{\mr B}^{\mr{ on}}$, the cross section
$\sigma_{\mr B}^{\mr {off}}$
develops a tail.
Although this tail becomes heavily suppressed at high energies, it is not
vanishing~\footnote
{
In~\cite{muta}, with which we analytically agree, this is not seen from
the figures.
}.
The other tail phenomenon is due to the universal ISR and is much more
pronounced, but again weaker than e.g. is observed for the narrow
$Z$-resonance shape.
In the figure, the radiatively corrected on-shell cross section
$\sigma_{\mr T}^{\mr{on}}$ is not shown.
As a matter of fact, we mention that the relative differences between
$\sigma_{\mr B}^{\mr{on}}$ and $\sigma_{\mr B}^{\mr{off}}$, and between
$\sigma_{\mr T}^{\mr{on}}$ and $\sigma_{\mr T}^{\mr{off}}$ are quite similar.

\setcounter{figure}{3}
\begin{figure}[bthp]
\begin{center}
\mbox{
\epsfysize=9.cm
\epsffile[0 0 530 530]{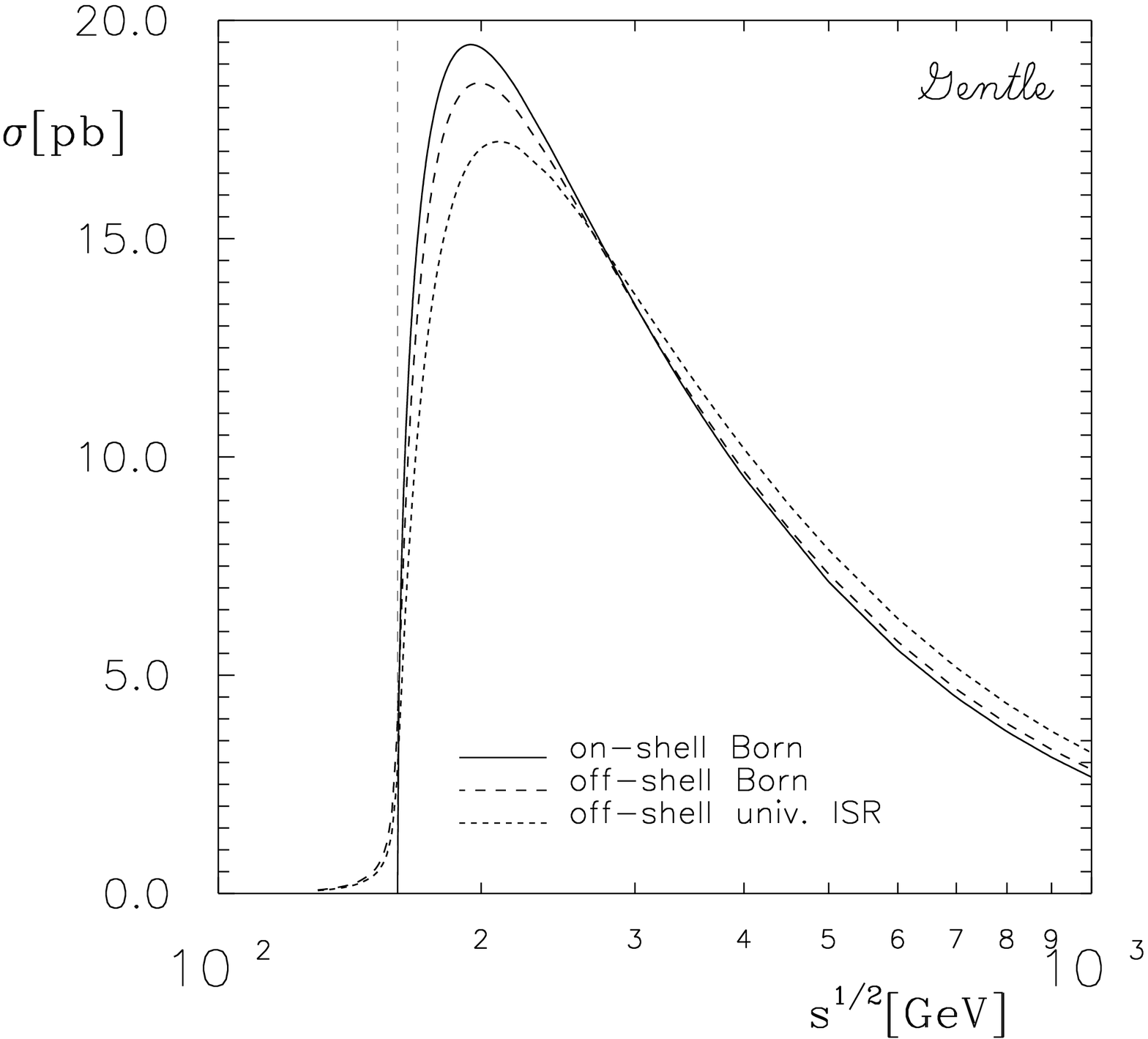}
}
\end{center}
\caption[]
{ \it
Total cross section for $W$-pair production:
$\sigma_{\mr B}$ and universal ISR corrections.
}
\label{fig4}
\end{figure}
%

\begin{figure}[thp]
\begin{center}
\mbox{
\epsfysize=9.cm
\epsffile[0 0 530 530]{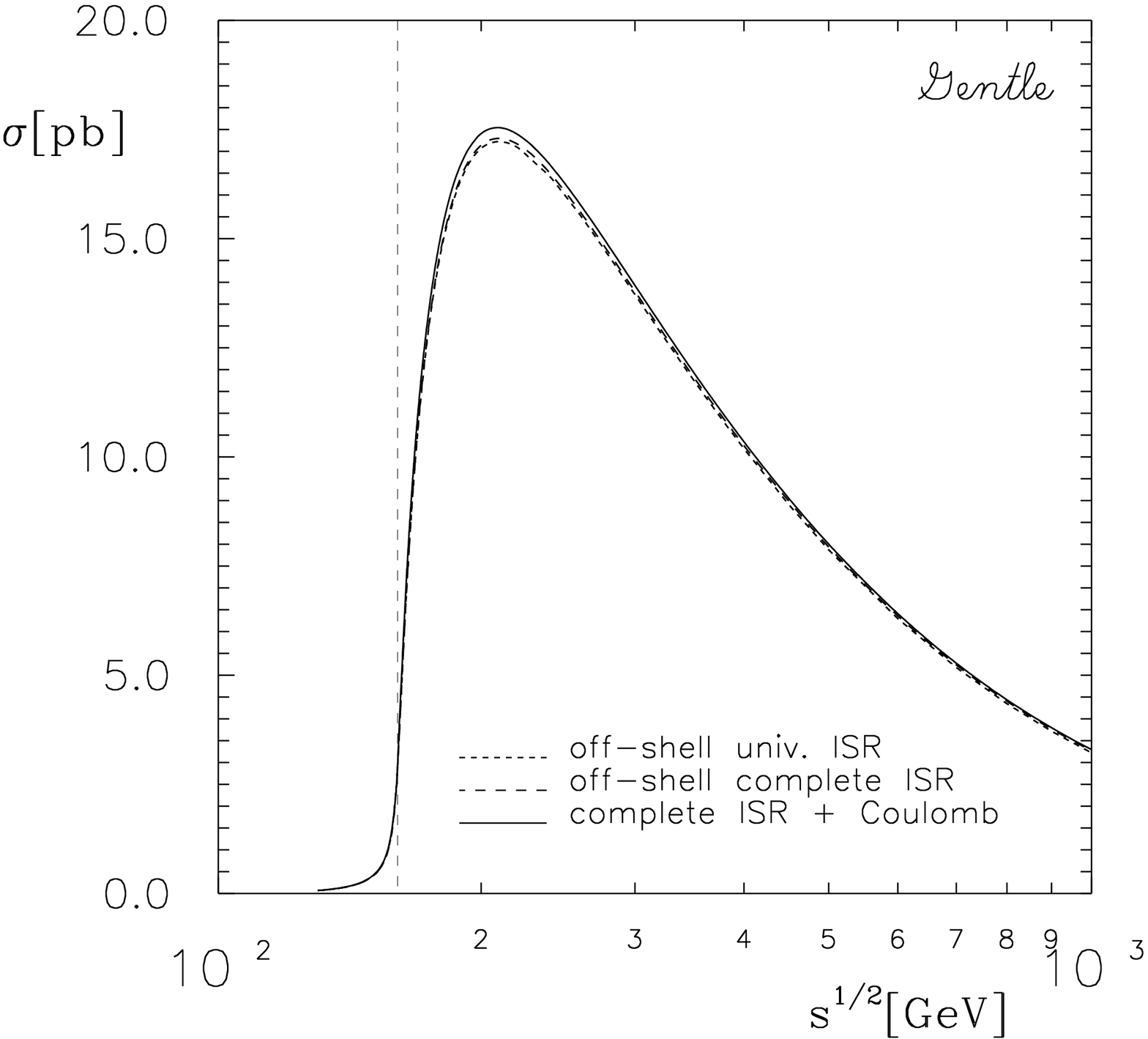}
}
\end{center}
\caption[]
{ \it
Total cross section for $W$-pair production with
ISR and Coulomb corrections.
\label{fig5}
}
\end{figure}

Figure~\ref{fig5} and table~\ref{tab1}
allow to estimate the differences between
the simple universal ISR approximation and the exact
treatment of ISR, which additionally includes the nonuniversal cross-section
contributions connected with the {\tt crab} diagram, and also the Coulomb
singularity.
The nonuniversal corrections $\sigma_{\hat S}$ and $\sigma_{\hat H}$
increase the
total cross section by
+0.4\%; +0.8\%; +1.5\% at $\sqrt{s}$ = 165$\div$190; 500; 1000
GeV, respectively. We quote these values for the on-shell case; the
off-shell corrections (in percent) are nearly the same.

\begin{table}[thbp]
\begin{center}
\begin{tabular}{||c|c|c|c|c|c||}
\hline\hline
 & & & & & \\
${\hat S}$ & ${\hat H}$ & $U_2$ & $C$ & $\sigma$, pb &
$\langle E_{\mr{ rad}} \rangle$, GeV
\\
 & & & & &
\\ \hline\hline
0 & 0 & 0 & 0 & 16.405 & 2.091
\\ \cline{5-6}
  &   &   &   & 13.634 & 1.158
\\ \hline\hline
0 & 0 & 1 & 0 & 16.401 & 2.153
\\ \cline{5-6}
  &   &   &   & 13.602 & 1.190
\\ \hline
1 & 0 & 0 & 0 & 16.471 & 2.091
\\ \cline{5-6}
  &   &   &   & 13.686 & 1.159
\\ \hline
0 & 1 & 0 & 0 & 16.405 & 2.090
\\ \cline{5-6}
  &   &   &   & 13.634 & 1.158
\\ \hline
1 & 1 & 0 & 0 & 16.470 & 2.091
\\ \cline{5-6}
  &   &   &   & 13.686 & 1.159
\\ \hline
0 & 0 & 0 & 1 & 16.783 & 2.107
\\ \cline{5-6}
  &   &   &   & 14.042 & 1.170
\\ \hline\hline
1 & 1 & 1 & 1 & 16.846 & 2.170
\\ \cline{5-6}
  &   &   &   & 14.063 & 1.201
\\ \hline\hline
\end{tabular}
\caption[]
{\it
The off-shell $W$-pair production cross section and the radiative energy
loss at two
energies: $E_{\mr{ beam}}$=95 GeV (upper rows), 88 GeV (lower rows).
The $\hat S$ and $\hat H$ switch on and off the soft and hard parts of
the non-universal QED corrections, the $U_2$ the second order
corrections, and $C$ the Coulomb correction.
\label{tab1}
}
\end{center}
\end{table}

On top of that, the Coulomb singularity yields a positive correction, which has
its maximum value of about 6\% at the threshold and asymptotically approaches
the on-mass-shell value of $\frac{1}{2}\pi\alpha=1.15\%$.
At large $\sqrt{s}$, in contrast to the threshold region, it is only
one of many \oalf\ corrections.
We compared numerically our equation~(\ref{barbeta}) with the predictions
of equations~(1) and~(9)~\footnote
{In equation~(9) of~\cite{khoze2},
we treated the arctg as defined between 0 and $\pi$.
}
of~\cite{khoze2},
and got very good agreement in the threshold region.
The predictions
of the non-relativistic calculation are slightly smaller, but the
absolute differences of the two calculations
do not exceed 0.5\%.
%
\subsection{The radiative energy loss}
%
Two potential            methods for a determination of $M_W$
are the direct reconstruction of events and the
measurement of the (upper and lower) energy end points in leptonic
$W$-boson decays.
They both rely on the knowledge of the {\em effective} beam energy, which
deviates from the beam energy itself by the radiative energy loss
$E_{\mr{rad}}$.
Its average value is
\ba
\langle E_{\mr{rad}}\rangle &=&
\frac{1}{\sigma_T(s)} \int ds' k_0 \int ds_1ds_2 \frac{d^3\sigma}{ds'ds_1ds_2},
\\
k_0
&=&
E_{n\gamma} = \frac{\sqrt{s}}{2} \left(1 - \frac{s'}{s} \right) .
\label{loss}
\ea
Since the radiative energy loss is essentially
due to ISR, with a possible smaller amount from initial-final interferences,
one may assume that the formulae of the foregoing sections cover the bulk of
effects.

Numerical estimates for the cross section  and the radiative energy loss are
collected in table~1 for two typical energies at LEP~200.
The radiative energy loss amounts to about 2200~MeV at $\sqrt{s}$=190 GeV
(1200~MeV at $\sqrt{s}$=176 GeV).
These shifts are mainly due to the universal, \oalf\ ISR corrections with
soft-photon exponentiation.
Further, there are at 190~GeV shifts of
+62; $\pm$0; +16 MeV due to: universal, non-exponentiated higher orders;
nonuniversal corrections; the Coulomb singularity (at 176~GeV:
+32; $\pm$0; +12~MeV, correspondingly).
Apparently, these corrections have to be taken into account properly, if one
aims at an accuracy of 50 to 100 MeV for the $W$-boson mass.

\bigskip

The numerical results of~\cite{169} were based on the corrected version
of the Fortran program.
\subsection{Conclusions}
We have performed the first complete, gauge-invariant calculation of
initial-state radiation corrections to off-shell $W$-pair production.
In the threshold region, we also include the Coulomb singularity.
The net improvements compared to the well-known universal ISR corrections are
relatively small, but non-negligible, at \ltwo.
They become more and more important at higher energies, and the final-state
radiation seems to be much less pronounced after the inclusion of the
auxiliary current into the ISR.

%
\section*{Acknowledgements}
We would like to thank
W.~Beenakker, A.~Denner, F.~Jegerlehner, W.~Hollik, V.~Khoze, R.~Kleiss,
G.~van Oldenborgh, B.~Pietrzyk for discussions.
D.B. and T.R. gratefully acknowledge the hospitality at the MPI M\"unchen,
where part of the work has been performed.
\\
We should like to thank D.~Lehner who helped us finding a mistake in
one of our {\tt SCHOON\-SCHIP} codes for the calculation of the
nonuniversal corrections.


\end{document}